# Field-free switching of perpendicular magnetic elements by using two orthogonal sub nanosecond spin orbit torque pulses

D. Suess[1,2], C. Abert[1,2], S. Zeilinger[1], F. Bruckner[1], S. Koraltan[1,2]

[1] Faculty of Physics, University of Vienna, 1090 Vienna, Austria

[2] Platform MMM Mathematics–Magnetism–Materials, University of Vienna, 1090

**Abstract:** *We propose a field-free switching mechanism that utilizes two spatially orthogonal spin-orbit torque (SOT) currents. Initially applied simultaneously, one of the currents is subsequently switched off. The superposition of these two currents results in an in-plane magnetization, which is not orthogonal to the remaining SOT current after the second one is deactivated. This symmetry-breaking procedure leads to reproducible and rapid switching, with field pulse durations as short as 0.25 nanoseconds.*

Magnetoresistive Random Access Memory (MRAM) has emerged as a promising candidate for a non-volatile memory cell due to its unique combination of speed and endurance. To enhance the durability of MRAM elements by separating the read-back current from the write current, Spin-Orbit Torque (SOT) switching is employed. However, SOT switching of out of plane devices require the breaking of symmetry to achieve reliable and deterministic switching, which was originally realized by applying an external field parallel to the SOT current direction [1]. Since then, various methods to circumvent the need for an additional bias field have been proposed, as reviewed by Krizakova et al. [2]. It includes using lateral geometry asymmetries [3], thickness asymmetries [4], [5], tilted anisotropy axis [6], in-plane magnets [7], exchange bias [8], combined SOT and STT [9], [10], crystal symmetries [11]. Sverdlov et al. [12] used two SOT pulses for switching. However, magnetic field-free switching could only be achieved if a geometric overlap of about 30% of the second pulse wire with the free layer is achieved, which is technologically very difficult to produce. In addition, the "write pulse 1" is applied before the second consecutive current, which requires a very precise timing of the pulses.

In this letter we propose a switching scheme that (i) does not require geometrical overlap of two wires (ii) does not require precise timing of the two pulses. To demonstrate and show the concept, we perform micromagnetic simulations including SOT through a damping ($H_{dl}$) and field-like ($H_{fl}$) torque term augmented to the Gilbert equation,

$$\partial_t \mathbf{m} = -\gamma \mathbf{m} \times \left(\mathbf{H}^{\text{eff}} - H_{\text{dl}} \mathbf{m} \times \mathbf{p} - H_{\text{fl}} \mathbf{p}\right) + \alpha \mathbf{m} \times \partial_t \mathbf{m} \tag{1}$$

The damping [13], [14] and field-like fields [15] are defined as:

$$H_{\text{dl}} = \frac{j_e \hbar}{2 e \mu_0 t M_s} \cdot \eta_{\text{dl}} \tag{2}$$

$$H_{\text{fl}} = \frac{j_e \hbar}{2 e \mu_0 t M_s} \cdot \eta_{\text{fl}} \tag{3}$$

where $j_e$ is the applied current density, $e$ the electron charge, $t$ the thickness of the ferromagnetic layer where the SOT acts on, $M_s$ the saturation magnetization of the ferromagnetic layer [16]. $\eta_{\text{dl}}$ and $\eta_{\text{fl}}$ are the damping and field-like efficiencies, respectively.

The concept presented in this letter utilizes two orthogonal SOT current pulses in cross bar structure as shown in Figure 1 (a-d). Figure 1 (a) shows the initial state in the free layer, which is the down state. Figure 1 (b) shows the first step, where currents are applied along the *x*-bar and *y*-bar. Consequently, in the region beneath the free layer (Free), the total current is the vectorial sum of these two individual currents. These currents are selected to be sufficiently strong, ensuring that they exceed the critical current required to switch the magnetization to an in-plane orientation [17]. When the currents surpass this critical threshold, the equilibrium magnetization aligns in-plane and orthogonally to the total current direction. The in-plane orientation is thus determined by the relative strengths of the $j_x$ and $j_y$ currents. For instance, if $j_y$ is significantly larger than $j_x$, the magnetization will predominantly orient in a direction close to the *x*-axis. If $j_y \sim j_x$ the magnetization will point at an angle of -45° with respect to the *x*-axis. Figure 1 (c) shows the second step, where the $j_y$ current is turned off. If the remaining $j_x$ current is less than the critical current needed for in-plane switching, the magnetization will rotate back towards the *z*-direction. This rotation is facilitated by the initial $m_x$ component of the magnetization, acting similarly to the application of an $H_x$ field, which is usually required for deterministic Spin-Orbit Torque (SOT) switching. Figure 1 (d) shows that the magnetization finally is reversed to the up state, when also the $j_x$ current is switched of. So the entire switching process from down to up is finished. The read out of the free layer can be done by a standard tunnel junction with a layer stack as shown in Figure 1 (d).

In order to systematically design the field-free switching concept, we investigate the final equilibrium magnetization as function of damping constant $\alpha$, field-like and damping-like fields as shown in Figure 2. Here, only a current $j_x$ is applied. The final states of the magnetization ($m_x, m_y, m_z$) state are shown. The SOT current that generates these fields is applied for 20 ns. In all simulations we neglect the effect of Joule heating, which would reduce the effective anisotropy in experiments for longer pulses due to the rise of temperature. The simulations are performed as single spin simulations. The magnetic parameter for the magnetic layer are: saturation magnetization $\mu_0 M_s$ = 1 T, effective anisotropy taking into account for crystalline anisotropy and shape anisotropy $K_{u,eff}$ = $10^5$ J/m³, effective easy axis **k** = (0,0,1) and film thickness $t$ = 1nm.

The first row shows, that with increasing damping-like torque field and zero field-like torque field the $m_x$ magnetization linearly increases with $H_{dl}$. At the critical field of $H_{dl}$ = 0.5 x $H_k$ the magnetization $m_x$ = 0, which indicates that here the magnetization rotates in-plane parallel to the y-axis.

The second row in Figure 2 shows that $m_y$ linearly changes with $H_{fl}$ until the critical field $H_{fl}$ = 1.0 x $H_k$ is reached. Since the field-like field is equivalent to an external field it is clear from the Stoner Wohlfarth model that at $H_{fl}$ = 1.0 x $H_k$ the magnetization fully rotates in-plane.

In the white region in the third row ($m_z$ component) the SOT current is sufficiently strong to rotate the magnetization in-plane.

This critical field strength is required for the total current ($j_x$ + $j_y$) in step 1 of the switching process. In the region marked by the orange polygon no equilibrium state is obtained and the system is in a dynamic self-oscillation.

In Figure 3 the final state is plotted for different in-plane states, described by the angle $\varphi$, for the case of a pure current in *x*-direction. Here, $\varphi$ describes the angle between the x-axis and the magnetization. The final state is plotted if only the $j_x$ current is applied. The first row shows the case, when the initial state of the magnetization points parallel to the *x*-axis (limit for $j_y \gg j_x$). The different columns show results for different damping constants $\alpha$. Depending on the strength of the field-like and damping-like SOT fields, the final state of $m_z$ is shown as a color code. In the second row the case is shown for $\varphi = -45°$. This corresponds to a scenario where, in the first switching step, $j_x$ equals $j_y$. It is important to note that for both investigated angles $\varphi$, the phase plots are quite similar. This indicates that the switching concept is quite robust across various values of $\varphi$, implying a substantial robustness with respect to the initial strengths of the superimposed $j_x$ and $j_y$ currents.

Figure 4 presents a detailed depiction of the magnetization dynamics influenced by applied currents over a specific time. In subfigure (a), the applied $j_x$ and $j_y$ currents are shown by the red lines, whereas the $m_z$ component of the magnetization is illustrated in blue. During the time interval of 2 ns to 4 ns, the system finds itself in state "S2". At this juncture, the collective strength of $j_x$ and $j_y$ currents sums up to 8 TA/m². This amount of current is sufficiently strong to rotate the magnetization in plane as depicted in the phase plot given in subfigure (b), marked as "S2". As we progress into the time window between 4 ns and 6 ns, only the $j_x$ current remains active, exerting a reduced current strength of 2 TA/m². Considering an initial state that is aligned in-plane with an angle $\varphi \sim$ -45 degrees (a scenario visually represented at the top row under state "S2"), the phase plots indicate a transition towards a magnetic state characterized by a positive $m_z$ component. This prediction obtained by the phase plots of the single spin simulations are confirmed through micromagnetic simulations illustrated in subfigure (a), demonstrating that the magnetization states labeled "S3" and "S4" indeed assume an upward orientation. Our finite element code magnum.pi is used for the simulations [18]. It takes into account the strayfield and magnetization inhomogeneities accurately. The lateral mesh size is 2.5 nm. In the perpendicular direction one layer of finite elements is used.

By reversing the $j_Y$ current the switching into the downward direction can be realized as shown in Figure 5 (a). In order to test how fast the switching can be realized, different current pulse durations have been applied, see Figure 5 (b). The time period $dt$ denotes the $j_x$ current pulse duration. The duration of the $j_Y$ current pulse is half the time period $dt$. The risetime of all pulses is assumed to be 0.1 ns. It can be seen that successful switching can be obtained for a current pulse length of $dt$ = 0.25 ns.

To summarize, a deterministic SOT switching of perpendicular MRAM cells is demonstrated, that relies on the superposition of two SOT current pulses. By switching off one of the current pulses earlier than the other, a symmetry breaking can be achieved, leading to a robust mechanism for deterministic switching in the sub nanosecond regime. This switching mechanism can not only by applied to MRAM elements but also for magnetic sensors to switch and modulate an out of plane reference system[19]. Performing two measurements of the sensor response for the two polarizations of the reference system and subtracting the sensor signal allows to reduce the sensor offset.

The financial support of the FWF project I 4917, P 34671 and I 6068 is acknowledged.

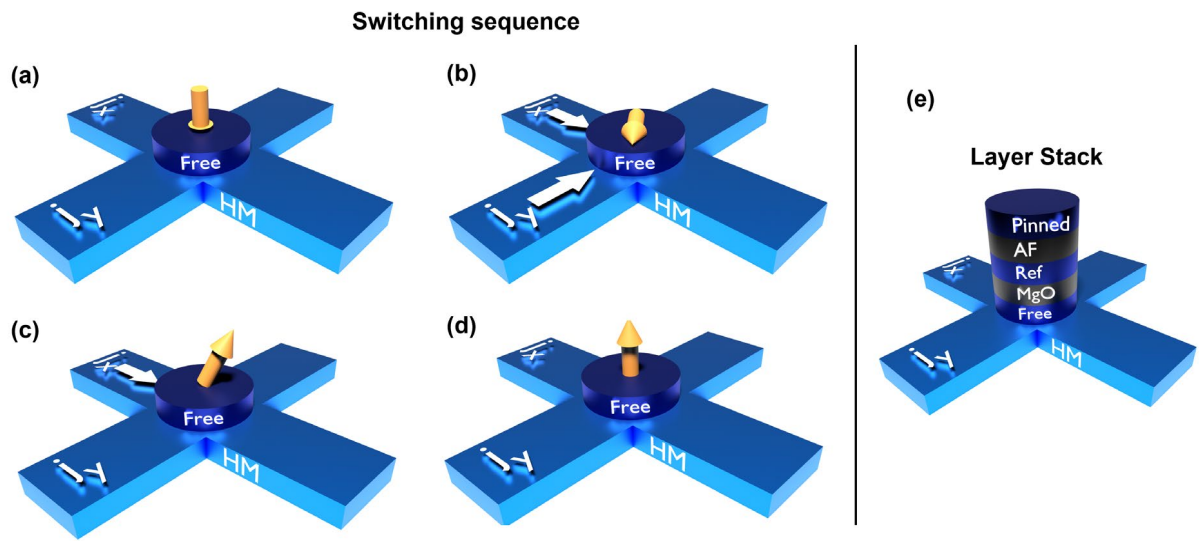

*Figure 1: (a-d) shows the switching sequence of a free layer (Free) for the different applied current strengths in a heavy metal (HM) cross bar structure. (a) initial state when no currents are applied (b) SOT currents in $j_x$ and $j_y$ are applied that switch the magnetization in plane. (c) the $j_y$ current is switched off so that only the $j_x$ current is applied that is smaller than the in-plane threshold current (d) final state when no currents are applied. (e) For the readout a standard tunnel junction with reference (Ref) and pinned layer (Pinned) can be used, that are antiparallel coupled via an AF-coupling layer (AF).*

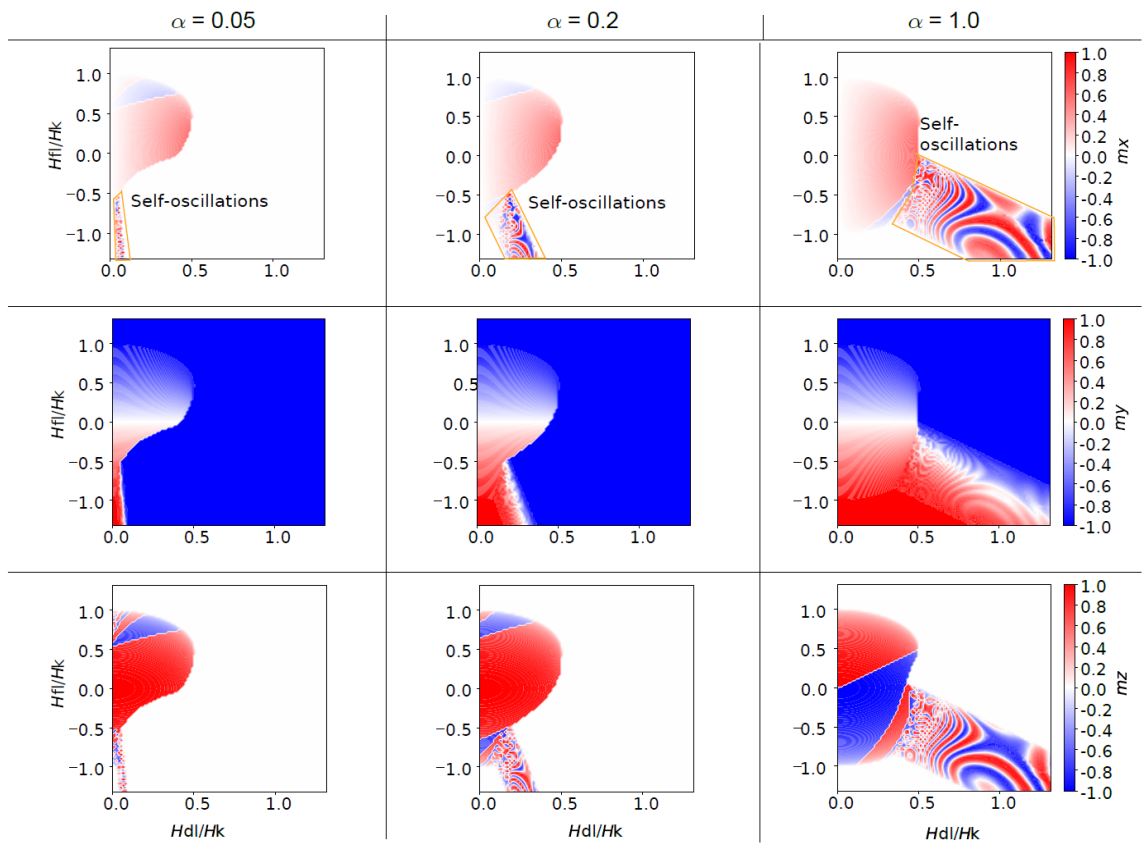

Figure 2: Phase diagrams of the equilibrium magnetization as function of damping and field-like fields
In the first, second and third row, the equilibrium $m_x$, $m_y$ and $m_z$ component are shown, respectively. The different rows are for different damping constants $\alpha$. The fields are given in units of the effective anisotropy field. The white regions in the third row show the regions where the critical current is exceeded rotating the magnetization in-plane.

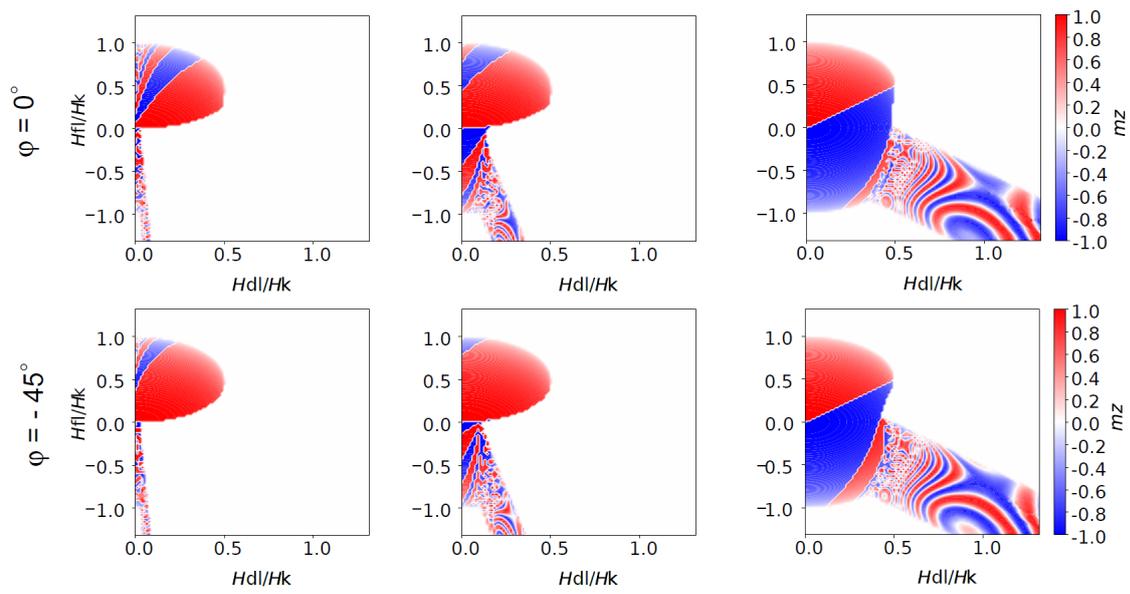

Figure 3: Final $m_z$ states for different in-plane initial states. The in-plane angle $\varphi$ is the angle between $x$-axis and magnetization. A SOT current in $j_x$ direction is applied.

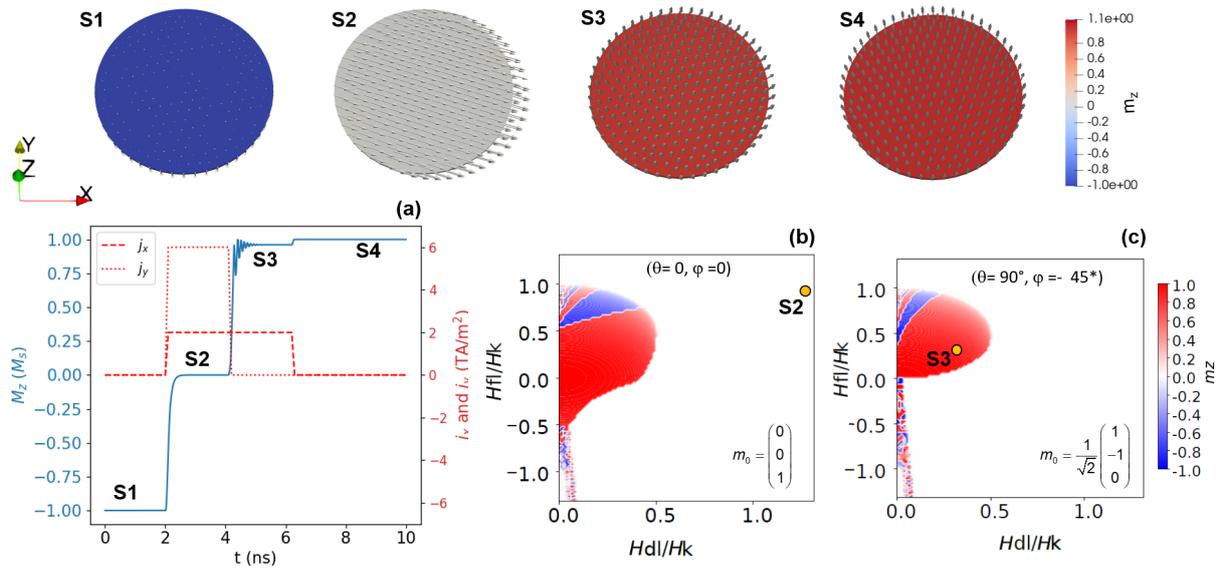

Figure 4: (a) Current sequences to switching the layer with SOT currents from down to up. The free layer is simulated with a finite element micromagnetic code. Parameters: α = 0.05, μ₀ $M_s$ = 1 T, A = 13e-12, $K_u$ = 10$^5$ + ½ μ₀ $M_s^2$ (J/m³), **k** = (0,0,1), $\eta_{dl}$ = 0.1, $\eta_{fl}$ = 0.07. r = 20 nm, t = 1nm. Between the time 2 and 4 ns the system is in the state "S2". The sum of $j_x$ and $j_y$ has a total strength of 8 TA/m² that leads to the field-like and damping-like torque terms has shown in subfigure (b) market with "S2". Here the torques are sufficient strong to rotate the magnetization in-plane ($m_z$=0). Between time 4 and 6 ns only a $j_x$ current is applied that leads to the current strength of 2 TA/m². The resulting effective field are shown in subfigure (c) as "S3". Here for an initial state that is approximately in-plane with φ = -45° (see in the top row the state "S2") the phase plots shows that the resulting magnetic state will be in the +$m_z$ state. This is confirmed by the micromagnetic simulation (subfigure b) that shows that state "S3" and "S4" are pointing in the up state.

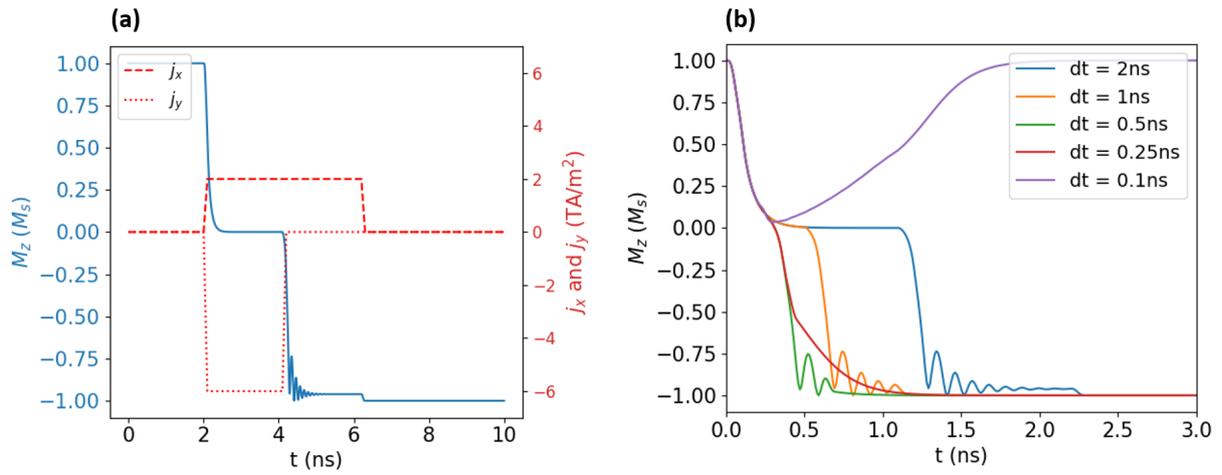

Figure 5: Current sequences to switching the layer with SOT currents from up to down. The subfigure (a) shows the applied $j_x$ and $j_y$ current (red lines) as well as the $m_z$ component of the magnetization (blue). Comparing with Figure 4a one sees that the polarity of the $j_y$ component determines the final polarity of the $m_z$ state. (b) The switching speed is evaluated. Here *dt* denotes the pulse length of the $j_x$ current pulse. The pulse length of the $j_y$ current pulse is about half of *dt*. The risetime of all current pulses is constant 0.1 ns. The time *t* on the x-axis in the right plot measures the time after the current pulses are applied.